\input{aipcheck}

\documentclass[
    ,final            % use final for the camera ready runs
  ]
  {aipproc}

\layoutstyle{6x9}

\usepackage{multicol}

\def\1{\c{c}}
\def\2{\c{C}}
\def\3{\.{I}}
\def\4{\"{a}}
\def\5{{\i}}
\def\6{$\beta$}
\def\7{\"{o}}
\def\8{\"{O}}
\def\9{\c{s}}
\def\0{\c{S}}
\def\*{\"{u}}
\def\,{\"{U}}
\def\;{\u{g}}
\def\:{\u{G}}
\begin{document}

\title{GeV Analysis of Mixed Morphology Supernova Remnants Interacting with Molecular Clouds}


\author{T\*l\*n Ergin}{
  address={Bo\;azi\1i University, Physics Department, 34342 Bebek, Istanbul/Turkey.}
}

\author{E. Nihal Ercan}{
  address={Bo\;azi\1i University, Physics Department, 34342 Bebek, Istanbul/Turkey.}
}

%\author{<author3>}{
% address={<common address for author2 and author3>}
%,altaddress={<author1 address>} % additional visiting address
%}

\begin{abstract}
The first remnants detected by the Fermi Gamma-ray Space Telescope were of the type of mixed-morphology supernova remnants interacting with molecular clouds. In this paper we are presenting preliminary results of the gamma-ray analysis of 4 selected mixed morphology remnants, G359.1-0.5, G337.8-0.1, G001.0-0.1, and G346.6-0.2, as well as  G349.7+0.2, in the 0.2 - 300 GeV energy range from the data collected by Fermi Gamma-ray Space Telescope for 3 years. G359.1-0.5, G337.8-0.1, and G349.7+0.2 were all detected with significances above 5 $\sigma$. The excess distribution of G359.1-0.5 shows extended gamma-ray emission, which coincides with the TeV gamma-ray source HESS J1745-303. G337.8-0.1 also shows an extended nature.
%Both remnants need to be studied in detail further to determine their extensions and spectral properties. 
 \end{abstract}

\footnote{Copyright 2012 American Institute of Physics. This article may be downloaded for personal use only. Any other use requires prior permission of the author and the American Institute of Physics.} {\footnotesize The following article appeared in T\*l\*n Ergin and E. Nihal Ercan, AIP Conf. Proc., 1505, pp. 265-268 (2012) and may be found at http://link.aip.org/link/?apc/1505/265 .}

 \maketitle

%%%%%%%%%%%%%%%%%%%%%%%%%%%%%%%%%%%%%%%%%%%%
%% MAINMATTER
%%%%%%%%%%%%%%%%%%%%%%%%%%%%%%%%%%%%%%%%%%%%

\section{Introduction}
The first supernova remnants (SNRs) detected by Fermi Gamma-ray Space Telescope's Large Area Telescope (Fermi-LAT), IC 443, W28, W51C, and W44, are middle aged mixed morphology (MM) SNRs interacting with molecular clouds. The interaction regions exhibit OH maser emission at 1720 MHz. 
%Mixed morphology (MM) remnants are predicted to have massive progenitors due to the fact that they exist in molecular cloud regions tracing star formation. 
Two criteria were used to select the SNRs for the analysis: First, from all the Green's Catalog \citep{GreensCatalog} SNRs that have not been associated with any of the 2nd year Fermi-LAT Catalog sources \cite{Nolan2012}, we selected the MM-type SNRs that  show evidence for interactions through maser detections. Second, SNRs that are intermediate or middle- aged remnants were chosen due to the fact that the estimated peak luminosity of gamma rays is reached at the early Sedov phase. As a cross-check of our analysis with the results in the literature \citep{Castro2010}, SNR G349.7+0.2 (not an MM-type SNR, but interacting with molecular clouds) was analyzed. Then the analysis continued with the MM-type SNRs. Here is a brief introduction to all of these sources:  

{\bf SNR G349.7+0.2:} The 3500 year old remnant, which is at a distance of 18.3-22.4 kpc, is a bright radio and X-ray emitter \citep{Lazendic2005}. The Chandra data revealed X-ray emission from a point source, CXOU J171801.0-372617, inside the SNR shell, with X-ray characteristics similar to the compact central object (CCO) class. Also five maser spots were detected in the direction of this SNR \cite{Frail1996}. G349.7+0.2 was also detected in GeV gamma rays as an unresolved point-like source positionally coinciding with the radio location of the SNR and showing a steepening spectral index above a few GeV \citep{Castro2010}. 	
 
{\bf SNR G359.1-0.5:}  The MM-type remnant at an age of at least 10000 years old and at a distance of 9.2 kpc, exhibits maser emission lines \citep{Yusef-Zadeh1995}. It was detected in radio, X-rays \citep{Ohnishi2011}, and in TeV gamma rays by H.E.S.S. (HESS J1745-303) \citep{Aharonian2008}. Recently, Hui et al. \citep{Hui2011} studied the GeV gamma-ray emission from HESS J1745-303 using 2 years of Fermi-LAT data. They found that a simple power-law was sufficient to describe the GeV spectrum with a spectral index of -2.6 at energies higher than 1 GeV, which could be connected to the spectrum of Region A of HESS J1745-303 in 1-10 TeV without any spectral break.

{\bf SNR G337.8-0.1:} The 12000 years old remnant's distance is estimated between 9 and 12.4 kpc \citep{Kothes2007}. It was detected as a thermal X-ray source by XMM-Newton \citep{Combi2008}. The MM-type remnant is also interacting with molecular clouds \citep{Koralesky1998}. In gamma rays it's associated with an EGRET source 3EG J1639-4702 \citep{Torres2003}.

{\bf SNR G346.6-0.2:} The 11000 years old remnant at a distance of about 5.5-11kpc is a mixed-morphology remnant \citep{Kothes2007}. Five masers were detected toward this SNR and they are all located along the southern edge of the remnant \citep{Koralesky1998}. It is also detected in X-rays by Suzaku \cite{Sezer2011}.

{\bf SNR G001.0-0.1:} This SNR is a MM-type 11000 year old remnant at a  distance of 8 kpc. Radio and maser emission exists, showing an interaction of the remnant with the surrounding ISM \citep{Hewitt2008}. Suzaku data reveals X-ray emission from the Sagittarius D HII region, where this SNR is located \citep{Sawada2009}. 
%TeV data was taken by HEGRA. 
	
\section{Observations and Analysis}

%\subsection{<A subsection>}

%\subsubsection{Observations}

Fermi-LAT is a pair conversion detector with a wide field of view  ($\sim$ 2.4 steradian) surveying the whole sky since 2008. LAT covers a broad energy range from 20 MeV to 300 GeV with an improved point-spread function (PSF) of 0.6$^{\circ}$ to 0.9$^{\circ}$ at 1 GeV and 0.11$^{\circ}$ to 0.15$^{\circ}$ at 10 GeV \cite{Atwood2009}. 
%The PSF is determined at lower energies by multiple scattering in the tracker. 
%At higher energies, the PSF approaches to the limit given by the granularity of the tracker channels (0.16$^{\circ}$-0.26$^{\circ}$ at 5 GeV and 0.11$^{\circ}$-0.15$^{\circ}$ at 10 GeV).
%LAT has a wide field of view (~2.4 sr), one-fifth of the entire sky. 
%Observations are done in the sky-survey mode and the viewing direction of Fermi changes every 1.5 hrs to achieve nearly uniform coverage of the full sky every two orbits. 
%To identify and study sources accurately, Fermi-LAT must be able to measure the locations of bright sources to within 1' (~0.02$^{\circ}$). 
%Fermi-LAT will have high sensitivity above 10 GeV, because there aren't many known types of sources at these energies that could be a potential background in the point-source analysis.

%\subsubsection{Analysis}
Fermi-LAT data was analyzed using Fermi Science Tools (FST) version v9r27p1. 
%The Earth limb is a very bright source of gamma rays for Fermi-LAT, and during observations it comes very close to the edge of the field of view. 
To remove the contribution from the gamma rays from the Earth's limb, which comes very close to the edge of the field of view during the observations, we cut out the gamma rays with reconstructed zenith angles greater than 105$^{\circ}$. We selected class 2 events in the energy range of 0.2-300 GeV. 
%At energies below 200 MeV and above 50 GeV, the contribution of the background gamma rays becomes comparable to the gamma rays coming from the SNRs. So, we will only include gamma rays above 200 MeV.
For all the remnants the gamma rays were taken from the circular region of interest (ROI) of radius 12$^{\circ}$ centered around the radio position of each SNR from the Green's Catalog \citep{GreensCatalog}. The analyzed area for each SNR was 2$^{\circ}$ $\times$ 2$^{\circ}$ centered at the same location as the ROI. The generated exposure maps had 600 $\times$ 600 pixels with 0.05$^{\circ}$$\times$ 0.05$^{\circ}$ bins. The gamma-ray events in each SNR data set were binned in energy at 15 logarithmic steps starting from 200 MeV and going up to 300 GeV. The IRF version used in the analysis was P7SOURCE$_{-}$V6. 
The typical model of an analysis region contains the SNR of interest and all (point-like and diffuse) background sources. The standard diffuse background model has two components: diffuse galactic emission (gal$_{-}$2yearp7v6$_{-}$v0.fits) and isotropic component (iso$_{-}$p7v6source.txt), which is a sum of the extragalactic background, unresolved sources, and instrumental background, where it's distribution is assumed to be isotropic. The point-like background sources were taken from the 2nd Fermi-LAT source catalog. Among the point-sources in the analysis region, only sources with significance values higher than 5 $\sigma$  are selected for the analysis. The background and source modeling was done by the binned likelihood analysis using gtlike of FST. 
All SNRs were fit as point-like sources having power-law (PL) type spectra. But to see how much contribution is made by the SNR, the residual maps like in Figure \ref{fig:2} are produced by subtracting the model counts map generated by the background model, which is not including the SNR, from the true counts map in the same region. Since the goal of this analysis was primarily to detect these SNRs and obtain preliminary results, we left the in-depth extension and spectral studies of these SNRs to be completed in future papers.
%\cite{Brown2000,BrownAustin:2000}. 
%\cite{Mittelbach/Schoepf:1990} a lacrimis? 
%\cite{Wang} 

\section{Results and Conclusion}
$~$
%%%%%%%%%%%%%%%%%%%%%%%%%%%%%%%%%%%%%%%%%%%%
%% Sample figure:
%%
%% The option [height=...] scales the picture to the given height,
%% without it it would be printed at its nominal size
%%%%%%%%%%%%%%%%%%%%%%%%%%%%%%%%%%%%%%%%%%%%

{\bf SNR G349.7+0.2: } The data was analyzed between 2008-08-04 and 2012-03-06. This remnant was detected at 10 $\sigma$, which is in agreement with the previous results in the literature \citep{Castro2010}.  The current results about this SNR are shown in Table \ref{tab:a}.
%\begin{figure}
%  \includegraphics[height=.25\textheight]{pics/g349_CompResid_2x2_4_MW_bw}
%  \caption{Comparison between residual maps for the model including SNR G349.7+0.2 (left panel) and the one without including SNR G349.7+0.2 (right panel). Both maps are smoothed with Gaussian kernel of 3.}
%  \label{fig:1}	
%\end{figure}

\begin{figure}
  \includegraphics[height=.27\textheight]{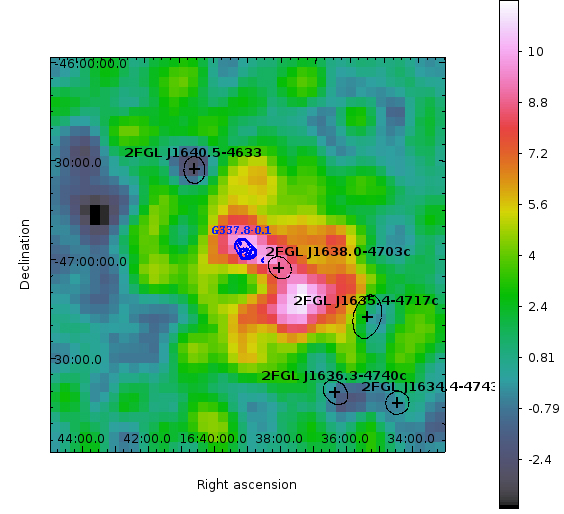}
  \includegraphics[height=.27\textheight]{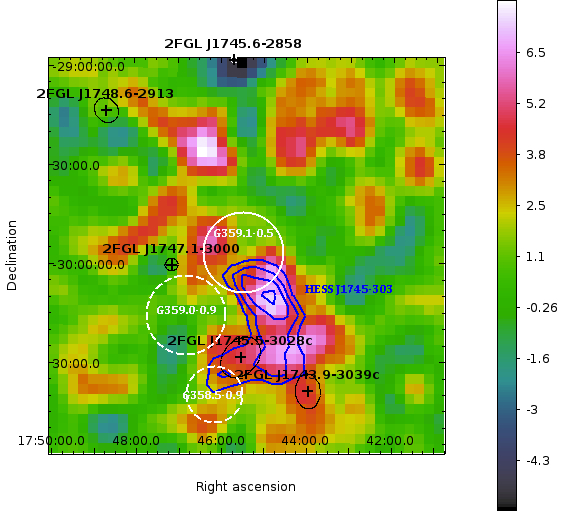}
  \caption{GeV residual maps of SNRs G337.8-0.1 (left panel) and G359.1-0.5 (right panel) are smoothed with a Gaussian kernel of 3. The background point sources are shown in black plus-markers and error circles. The remnants are shown with solid or dashed lines. On the left panel radio continuum data is shown in blue contours (0.2-0.85 Jy). On the right panel the blue contours represent the significance levels (5-7.8$\sigma$) for HESS J1745-303.}
  \label{fig:2}	
\end{figure}

{\bf SNR G359.1-0.5:} The data was analyzed between 2008-08-04 and 2012-06-26. Initially, the data was analyzed only adding the SNR as an extra source to the model, because it didn't exist in the 2nd Fermi Catalog.  To see the excess from the SNR, the residual map shown in the right panel of Figure \ref{fig:2} was produced with a model file, which didn't have the SNR included. There is a partial overlap between the HESS J1745-303 significance contours and the excess regions in this map. The background model was modified by adding the 3 hot spots at locations. The locations were RA1=266.608$^{\circ}$ Dec1=$-$29.44$^{\circ}$, RA2=266.203$^{\circ}$ Dec2=$-$30.19$^{\circ}$, and RA3=266.100$^{\circ}$ Dec3=$-$30.44$^{\circ}$ for hot spot 1, 2 and 3, respectively. Only spot 3, which is inside the SNRs radio shell, was detected with a significance of bigger than 5 $\sigma$. Therefore, we assumed spot 3 to be a signal from the intersection region of G359.1-0.5 with HESS J1745-303. To resolve this region and to understand the origin of the GeV and TeV gamma rays, this data needs to be further investigated.

%\begin{figure}
%  \includegraphics[height=.25\textheight]{pics/g359_CMAP_Resid_G359_zoom}
%  \caption{Residual map for the model not including SNR 359.1-0.5, which is smoothed with a Gaussian with a kernel of 3. The background point sources are shown in black with yellow error circles. The remnant is shown with blue circle. Also H.E.S.S. source HESS J1745-303 is given in blue color. The weak sources 2FGL J1745.5-3028c, 2FGL J1743.9-3039c had significance values smaller than 5$\sigma$ and they were not taken into the model.}
%  \label{fig:3}	
%\end{figure}
{\bf SNR G337.8-0.1:} The data between 2008-08-04 and 2012-06-26 was analyzed. The closest 2nd Fermi Catalog source to G337.8-0.1 is 2FGL J1638.0-4703c at a distance of 0.17$^{\circ}$ away from the ROI center and its detection significance in the catalog was given as $\sim$ 14 $\sigma$. The residual map on the left panel of Figure \ref{fig:2}, however, shows that 2FGL J1638.0-4703c is located between the SNR's radio position, which has an excess of gamma rays of 6$\sigma$, and another unknown gamma-ray source that has a significance of 12$\sigma$ at RA=249.34$^{\circ}$ and Dec=$-$47.205$^{\circ}$. This shows that the gamma-ray excess is extended and needs to be further investigated.

{\bf SNR G001.0-0.1 and SNR G346.6-0.2:} The data for these two remnants were analyzed in the time range between 2008-08-04 and 2012-02-29, but no gamma-ray excess at the positions of the radio detected remnants. 
%\begin{enumerate}
%\item 
%\end{enumerate}

%%%%%%%%%%%%%%%%%%%%%%%%%%%%%%%%%%%%%%%%%%%%
%% SAMPLE TABLE
%%
%% Shows the use of \tablehead and \tablenote
%% macros
%%%%%%%%%%%%%%%%%%%%%%%%%%%%%%%%%%%%%%%%%%%%

\begin{table}
\begin{tabular}{cccccccc}
\hline
      \tablehead{1}{c}{b}{Name\\ $~$}
  & \tablehead{1}{c}{b}{RA\\ $\left[^{\circ} \right]$}
  & \tablehead{1}{c}{b}{Dec\\$\left[^{\circ}\right]$}
  & \tablehead{1}{c}{b}{Significance\\$\left[ \sigma \right]$}
  & \tablehead{1}{c}{b}{Spectral \\ Index}
  & \tablehead{1}{c}{b}{Total Flux\\ $~~~$$\left[ 10^{-8}~ph~cm^{-2}~s^{-1} \right]$}   \\
\hline
G359.1-0.5  & 266.100 & -30.44  & 6   & 2.588 & 4.452  \\
G337.8-0.1  & 249.750 & -46.98  & 6   & 2.517 & 5.194  \\
G349.7+0.2 & 259.530 & -37.43  & 10 & 2.444 & 5.786 \\
G346.6-0.2 &  257.575 & -40.18  &  -   &     -      &    -        \\
G001.0-0.1 &  267.125 & -28.15  &  -   &     -      &    -        \\
\hline
\end{tabular}
\caption{Results from the analysis of the 3 years of Fermi-LAT data in the energy range of 200 MeV and 300 GeV. }
\label{tab:a}
\end{table}

%\cite{EVH:Office} 

%\begin{enumerate}
%\item
%Another item with sub entries
%\begin{enumerate}
%\item
%A sub entry \cite{Wang}
%\item
%Second sub entry
%\end{enumerate}
%\item
%The final item with normal label.
%\end{enumerate}

%\begin{description}
%\item[Infandum]
% regina 
%\item[Sed]
% si tantus amor 
%\end{description}

%\cite{Liang:1983} 
% \cite{SJ:1999}

%\section{Conclusion}

%\cite{Knuth:WEB} 

%%%%%%%%%%%%%%%%%%%%%%%%%%%%%%%%%%%%%%%%%%%%%%%%
%% BACKMATTER
%%%%%%%%%%%%%%%%%%%%%%%%%%%%%%%%%%%%%%%%%%%%%%%%

\begin{theacknowledgments}
T. Ergin is thankful for the support by the postdoctoral fellowship of T\,B\3TAK (The Scientific and Technological Research Council of Turkey). Special thanks to Dr. Emma de \~{O}na Wilhelmi for providing me with the H.E.S.S. significance map. 
\end{theacknowledgments}



\bibliographystyle{aipproc}   % if natbib is available

%\bibliography{sample}

% \typeout{******************************************}
%  \typeout{** Please run "bibtex \jobname" to optain}
% \typeout{** the bibliography and then re-run LaTeX}
% \typeout{** twice to fix the references!}
% \typeout{******************************************}
% \typeout{}
 %}

%%%%%%%%%%%%%%%%%%%%%%%%%%%%%%%%%%%%%%%%%%%
%% The following lines show an example how to produce a bibliography
%% without the help of the BibTeX program. This could be used instead
%% of the above.
%%%%%%%%%%%%%%%%%%%%%%%%%%%%%%%%%%%%%%%%%%%
%\begin{multicols}{2}

%\end{multicols}
%\endinput

\end{document}